\documentclass[sigconf,10pt,nonacm]{acmart}

\setcopyright{none}
\acmConference[]{}{}{}
\acmBooktitle{}
\acmDOI{}
\acmISBN{}

\usepackage{amsmath}
\usepackage{array}
\usepackage{booktabs}
\usepackage{multirow}
\usepackage{subcaption}
\usepackage{graphicx}
\usepackage{xspace}
\usepackage{listings}
\usepackage{enumitem}
\usepackage{tikz}
\usepackage{makecell}
\usetikzlibrary{arrows.meta,positioning,shapes.geometric,shapes.misc,fit,calc,backgrounds}
\usepackage[nameinlink,noabbrev]{cleveref}

\newcommand{\system}{SysPilot\xspace}
\newcommand{\lpcl}{LPCL\xspace}

\newcommand{\vlcs}{VLCs\xspace}

\newcommand{\openmp}{OpenMP\xspace}
\newcommand{\openblas}{OpenBLAS\xspace}
\newcommand{\libtorch}{LibTorch\xspace}
\newcommand{\pytorch}{PyTorch\xspace}
\newcommand{\para}[1]{\vspace{2pt}\noindent\textbf{#1.}}

\AtBeginDocument{}

\begin{document}

\title{SysPilot: Cross-Library Performance Autotuning via Library Performance Contracts}

\author{Yineng Yan}
\email{yinengy@utexas.edu}
\affiliation{\institution{University of Texas at Austin}\city{Austin}\state{TX}\country{USA}}

\author{Haowei Chen}
\email{hwchen@cs.utexas.edu}
\affiliation{\institution{University of Texas at Austin}\city{Austin}\state{TX}\country{USA}}

\author{Christopher J. Rossbach}
\email{rossbach@cs.utexas.edu}
\affiliation{\institution{University of Texas at Austin}\city{Austin}\state{TX}\country{USA}}
\affiliation{\institution{Microsoft}\country{USA}}

\renewcommand{\shortauthors}{Yan, Chen, and Rossbach}

\begin{abstract}
Modern scientific and machine-learning applications compose many independently developed libraries. When these libraries run together, their independently managed runtimes can oversubscribe shared resources: using the same CPU cores, fighting over GPUs, or contending for memory bandwidth. Avoiding such contention requires careful configuration across library boundaries. Existing autotuners can search large configuration spaces, but they typically treat the target application or system as a black box. They do not directly investigate what resources are shared between libraries, which runtime signals correlate with interference, or how tuning actions should be coordinated across library boundaries. As a result, they are inefficient for finding optimal configurations to mitigate contention. 

This paper presents \system\footnote{SysPilot is open source at \url{https://github.com/yinengy/SysPilot}.}, a cross-library performance autotuning system based on \emph{Library Performance Contract (LPC)}. A Library Performance Contract captures a library's resource demands, assumptions, controls, and observability signals in \lpcl, a specification language for describing library resource requirements. \system synthesizes contracts from documentation and code artifacts, composes contracts from the libraries used by an application, derives resource-conflict hypotheses, profiles the application to validate those hypotheses, and selects candidate configurations. This design makes the Library Performance Contract a reusable unit of performance knowledge: a contract for a library can be reused across many applications that depend on that library. We evaluate \system on a benchmark suite of cross-library contention scenarios. The results show that \system matches the performance of autotuners that rely on application-specific manual information while retaining cross-application reuse. Compared with general-purpose autotuners such as OpenTuner and SMAC3, \system improves benchmark performance metrics by up to an additional 38\% over the best configuration found by these autotuners, while using 3$\times$ fewer runs.
\end{abstract}



\maketitle

\section{Introduction}

Modern high-performance applications are more often built by composing libraries rather than by implementing all computation and resource management in one codebase. For example, scientific and machine-learning workflows often combine numerical libraries, data-processing libraries, and accelerator runtimes,
such as NumPy, SciPy, CuPy, \pytorch, TensorFlow, \openblas, and Intel MKL~\cite{numpy,scipy,cupy,pytorch,tensorflow2015,openblas,mkl}.
This composition model allows applications to reuse optimized kernels and hardware-specific implementations. However, it also means that resource management is delegated to multiple independently designed library runtimes that may share and contend for different kinds of system resources.

CPU-parallel libraries such as \openmp~\cite{openmp}, \openblas~\cite{openblas}, and Intel MKL~\cite{mkl} create worker threads that consume CPU, cache, and memory. Data-processing libraries such as TensorFlow \texttt{tf.data}~\cite{tfdata}, Hugging Face \texttt{datasets}~\cite{hf_datasets}, and database systems~\cite{duckdb,postgres} create workers, queues, and prefetch buffers that consume memory, storage bandwidth, and network or object-store bandwidth. GPU libraries such as \pytorch~\cite{pytorch}, TensorFlow~\cite{tensorflow2015}, cuDNN~\cite{cudnn}, cuBLAS~\cite{cublas}, and DALI~\cite{dali} allocate device memory, launch kernels, and issue host-device transfers that consume GPU compute, device memory capacity, memory bandwidth, and PCIe or NVLink bandwidth. Each library may expose knobs for its own resource usage, such as thread counts, batch sizes, queue depths, memory limits, or communication buffer sizes. These controls are local: they can change how one runtime consumes CPU, GPU, memory, or network resources, but they do not capture how that resource use overlaps with other runtimes in the same application. As a result, a configuration that is efficient for one library in isolation can still cause contention when several libraries are sharing the same resources.

Recent work on Virtual Library Contexts (VLCs)~\cite{vlcs} identifies this as a fundamental challenge in library composition: many libraries assume they own all process resources or expose only process-level configuration, so composing them can over-allocate cores or devices and degrade performance. 
Mechanisms for controlling resources at library granularity are emerging. For example, \vlcs provide such a mechanism: they can virtualize libraries and partition resources without modifying library code. 
However, a mechanism does not by itself detect which libraries interfere, which shared resources are oversubscribed, or which resource partition yields the optimal performance. 
Cross-library contention mitigation still needs a diagnosis and tuning layer.

Traditional autotuners formulate tuning as search over an application-defined configuration space using measured performance feedback~\cite{opentuner,orio,ytopt,smac,spearmint,hyperopt,autoweka}. They are effective when the search space is compact or when enough trials are available, but they do not represent why independently managed libraries interfere. Recent LLM-assisted tuners~\cite{dbbert,gptuner,lambda_tune} move beyond purely black-box search by reading documentation, web pages, or code artifacts to generate tuning advice or candidate configurations. In \Cref{sec:motivation}, our motivating example shows that an LLM-assisted tuner given additional application/library descriptions could infer that the two tasks are \libtorch workloads competing for CPU cores and that the useful action is to search over asymmetric core partitions, rather than blindly exploring all per-core assignments. Such information greatly reduces the number of trials to find the optimal configuration by up to 6.4$\times$ compared with traditional autotuners.

However, existing LLM-assisted tuners still tend to organize knowledge around one target application or system~\cite{dbbert,gptuner,lambda_tune}. This is helpful for a single workload, but it misses an important opportunity for reuse. In the \libtorch example, the exact workload split is workload-specific, but the stable part of the problem belongs to the library: \libtorch creates library-managed worker parallelism, consumes shared CPU compute, and exposes runtime signals that can reveal contention. An application-specific manual, prompt, or search-space design must be rebuilt or reinterpreted for the next workload, even if that workload uses the same library and suffers from the same class of contention. Reusable library-level knowledge can instead transfer across applications that depend on the same library.

\system is a novel solution for cross-library performance tuning based on this observation. It treats the library as the first-class application building block and captures a library's performance characteristics using the \emph{Library Performance Contract Language} (\lpcl), a description language designed to capture the performance-relevant behavior of a library in a machine-readable form, including the resources the library uses, the controls that can change its behavior, and the signals that indicate performance degradation or contention.

Given \lpcl{} for the libraries used by an application, \system composes them to generate hypotheses about potential resource conflicts. Composition is a static analysis over contracts.
The key idea is that \lpcl gives each contract a common vocabulary for describing performance-relevant resources, controls, and signals. Once multiple contracts are expressed in this language, \system can compare them systematically and identify overlapping resource declarations across libraries. These hypotheses are not treated as proof of slowdown. Instead, they determine which signals to collect, such as context-switch rate and memory footprint. 
Only hypotheses supported by profiling evidence generate corresponding candidate configurations. This \emph{hypothesis-guided} workflow narrows the search space and ties each candidate configuration to an explicit diagnosis.

Organizing performance knowledge as library contracts makes that knowledge reusable: \lpcl written once for popular libraries like OpenMP and PyTorch can be applied across many applications that depend on those libraries, rather than being rebuilt from scratch for each target application. In this sense, \system shifts autotuning from an application-specific manual-building exercise to a cross-library reasoning problem over reusable library performance contracts.

An important aspect of this design is that LPCL is intended to be both \emph{authorable} and \emph{generatable}. A contract may be written directly by a library developer or performance expert, allowing human knowledge to be expressed in a structured, machine-readable form. To reduce the burden of manual authoring, \system also includes an LLM-assisted automatic generation pipeline that harvests artifacts such as documentation, websites, configuration interfaces, and source code, extracts evidence from them, and synthesizes an LPCL contract. This generation pipeline is an important practical mechanism for obtaining contracts at scale.

We make the following contributions:
\begin{itemize}[leftmargin=*]
  \item We describe \emph{Library Performance Contracts} as a reusable unit of performance knowledge for cross-library resource contention.
  \item We present \lpcl, a specification language that describes a library's performance characteristics, including the resource requirements, the controls that can change its resource usage, and the signals that indicate performance degradation.
  \item We design and implement \system, an autotuner that synthesizes and checks \lpcl, composes contracts into conflict hypotheses, profiles and collects signals, and searches for performant configurations.
  \item We build \texttt{LibContentionBench}, a benchmark suite that characterizes cross-library contention scenarios, and show that \system achieves up to a 38\% higher normalized score using 3$\times$ fewer trials than existing generic autotuners like OpenTuner and SMAC3 without writing one-off application-specific manuals to guide autotuning.
\end{itemize}

\section{Motivation}
\label{sec:motivation}
Composing libraries that share the same resources may lead to resource contention and performance degradation. For example, a C++ \libtorch hyperparameter tuning task for machine learning models can be accelerated by parallelizing runs to leverage the large memory capacity and massive parallelism \cite{hyperparameter,hpc_hyperparameter}. When running two tasks in parallel, they contend for the same CPU resources. Because each task has different resource requirements, the optimal resource configuration is asymmetric: e.g., in a 24-core, two-socket NUMA system, the optimal configuration may assign 6 cores to the light task and 18 cores to the heavier task. The core assignment should also follow the NUMA topology.

For a general-purpose autotuner, however, finding this configuration is not trivial. If the tuner is given only a resource-partitioning interface and end-to-end performance feedback, a natural encoding is to treat each of the 24 cores as an independent binary decision: assign this core to one of the tasks. This creates a $2^{24}$-point search space. This encoding hides the useful structure of the problem. The tuner is not really choosing among arbitrary per-core assignments; it is trying to find a good resource split between two competing library runtimes. A generic black-box autotuner does not know this distinction. It can only test candidate assignments and observe their performance, so it may spend many trials exploring configurations that are unlikely to resolve the underlying contention.

\begin{figure}[t]
\centering
  \includegraphics[width=\linewidth]{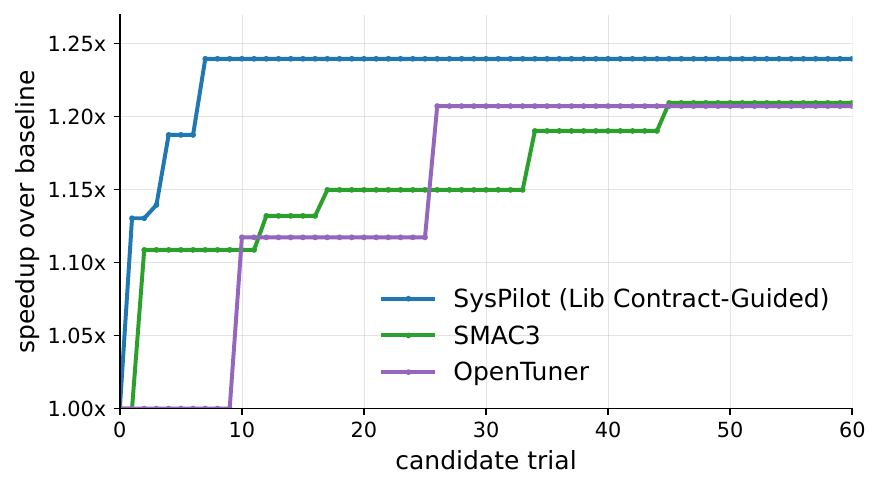}
\caption{
Best-so-far performance on the LibTorch hyperparameter tuning resource-partitioning example. The speedup is relative to the baseline that evenly splits cores between the two tasks.
\system finds the optimal configuration within 7 trials.
OpenTuner and SMAC3 require 26 and 45 trials, respectively, to reach near-optimal configurations. (Upper-left is better)
}
\label{fig:vlc_motivation}
\end{figure}

We use a simple experiment to illustrate how missing information has a direct effect on tuning efficiency. We implement a C++ \libtorch parallel hyperparameter tuning workflow on a transformer-based language model with 8 heads, 6 layers, and an embedding size of 512. The training dataset is {\tt wikitext2} \cite{wikitext}. We run two tuning tasks in parallel on a machine with two 12-core Intel E5-2650 v4 sockets, 128 GB of RAM. Each task can be configured with a variable number of CPU cores to avoid contention. We measure total training time for tuning a fixed number of hyperparameters.

Figure~\ref{fig:vlc_motivation} shows the performance of the best configuration found so far by different autotuners. We compare \system against two representative general-purpose autotuners: OpenTuner~\cite{opentuner}, an ensemble-based program autotuning framework that adaptively selects among multiple search techniques, and SMAC3~\cite{smac}, a sequential model-based algorithm configuration framework based on Bayesian optimization. Because OpenTuner and SMAC3 treat candidate configurations largely as opaque points in a search space, they require 26 and 45 trials, respectively, to reach near-oracle configurations. In contrast, \system uses library-level information to focus the search on contention-relevant resource partitions and reaches the best configuration within 7 candidate trials. The lesson is not merely that one optimizer is better than another. Rather, the example shows that the hard part is often not the numerical search procedure itself, but the absence of information that explains what resource is shared, which library controls affect that resource, and which class of actions is likely to relieve contention.
\section{Related Work}
\label{sec:related}

\para{Library resources and composition} The problem of composing parallel libraries has a long history. Lithe~\cite{lithe} proposes a common substrate that allows parallel runtimes to coordinate processors, but libraries must be ported to the Lithe interface. BOLT~\cite{bolt} reduces oversubscription for nested OpenMP parallelism by changing the OpenMP runtime. Lightweight Contexts~\cite{lwCs} divide a process into separate resource and protection domains, but require kernel support. \vlcs~\cite{vlcs} provide a user-space mechanism for library-level virtualization inside a single process and can partition resources across libraries without modifying the libraries or the operating system. \system is complementary to these mechanisms. It does not replace resource-management tools; it decides which tool or conventional library knob is likely to address a validated cross-library conflict.

\para{Autotuning and configuration search} Autotuners search configuration spaces using empirical feedback. OpenTuner~\cite{opentuner} provides an extensible ensemble of search techniques; Orio~\cite{orio} uses annotations to generate and tune optimized code variants; ytopt~\cite{ytopt} uses model-based optimization for scientific applications; and SMAC~\cite{smac}, Spearmint~\cite{spearmint}, Hyperopt~\cite{hyperopt}, and Auto-WEKA~\cite{autoweka} popularized Bayesian or model-based optimization for algorithm and hyperparameter configuration. FFTW~\cite{FFTW} uses empirical planning to select fast transform implementations. These systems are effective when the search space is well defined, but they generally treat the target application as a black box. \system instead uses a structured representation to shrink and organize the search space: it identifies plausible conflicts using \lpcl, validates them with runtime signals, and uses validated hypotheses to narrow the configuration search space.

Prior work on tuning database systems shows that optimizers can benefit from representations richer than raw performance measurements. OtterTune~\cite{ottertune} transfers tuning knowledge across database deployments; CDBTune~\cite{cdbtune} uses reinforcement learning for cloud database tuning; and DB-BERT~\cite{dbbert}, GPTuner~\cite{gptuner}, and \(\lambda\)-Tune~\cite{lambda_tune} use language models to extract or generate tuning guidance from textual artifacts. In networking, UNUM~\cite{unum} uses a Transformer-based state embedder to learn latent representations of network conditions and improve network controllers such as congestion control and adaptive bitrate streaming. These systems show that tuning and control can improve when the optimizer is given more informative representations than raw measurements alone. \system adopts this lesson, but changes the unit of representation and reuse. Existing database tuners typically organize knowledge around one DBMS, deployment, or workload instance, while UNUM learns a representation for network-control state. \system instead represents reusable library-level knowledge: a Library Performance Contract captures the resources, controls, signals, and validation rules of a library so that the same knowledge can transfer across applications that use that library.

\para{Instance-specific optimization} A broader line of systems work argues that performance mechanisms should be specialized to the data, workload, hardware, or operating context of a particular deployment. The vision for instance-optimized data systems~\cite{kraska_instance_optimized} argues that many data-system components, including query optimization, indexing, scheduling, and storage layout, can be specialized to a particular system instance. \system adopts the same principle for cross-library tuning: library contracts provide reusable, general knowledge about resources and controls, but profiling is required to determine which contract-derived hypotheses actually hold for the current application. Thus, \system combines reusable library-level knowledge with instance-specific validation to guide instance-specific tuning.

\section{Design}
\label{sec:design}

\begin{figure}[t]
  \centering
  \includegraphics[width=\linewidth]{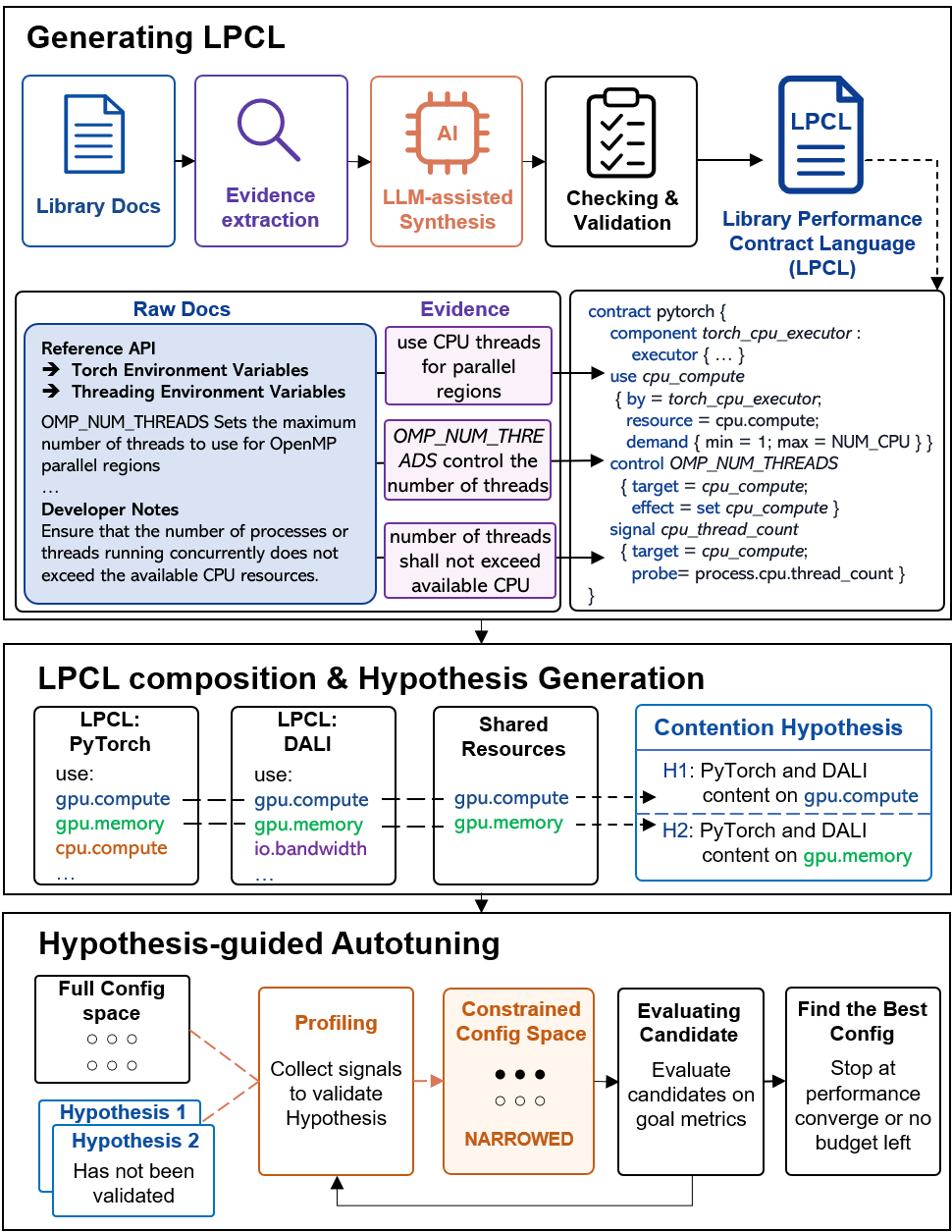}
  \caption{\system architecture.}
  \label{fig:architecture}
\end{figure}

\system is a cross-library autotuning framework that turns library-level performance knowledge into a reusable Library Performance Contract (LPC) and uses it to guide autotuning. \Cref{fig:architecture} shows the architecture of \system.

A Library Performance Contract describes how a library uses shared resources, which assumptions its runtime makes, which control knobs can change its resource demand and behavior, and which runtime signals indicate potential resource contention and performance degradation. Library developers or performance experts can write contracts in the domain-specific language \lpcl{} described in \Cref{sec:lpcl}. To reduce user burden, \system includes an LLM-assisted contract generation pipeline that turns library documentation, code, and API artifacts into \lpcl.
At runtime, \system composes the \lpcl for the libraries used by the application, derives contention hypotheses, profiles the workload to validate those hypotheses, and searches only over candidate configurations that are tied to validated conflicts.

\subsection{Library Performance Contracts}

A Library Performance Contract is a specification that characterizes performance-sensitive aspects of a library such as resource use, assumptions, control knobs, and performance metrics. A Library Performance Contract can be written in the Library Performance Contract Language (\lpcl), a domain-specific description language as a collection of declarations including:

\para{Components} 
Components are library-internal mechanisms that may consume or regulate resources. 
Examples include executors, queues, allocators, and backends.
Components are not shared resources by themselves. 
For example, an \openblas worker pool is an executor component; it becomes relevant to contention because it uses the shared resource \texttt{cpu.compute}.

\para{Uses} A use declaration states that a component consumes or constrains a shared resource. For example, a BLAS executor uses \texttt{cpu.compute} with a worker demand in \([1,\texttt{host.cpu\_slots}]\). Use declarations are the main input to contract composition because they make resource overlap explicit.

\para{Assumptions} Assumptions describe conditions under which the library's local resource policy is expected to behave well. 
Examples include \texttt{exclusive} for a workspace that assumes no competitors. Assumptions are useful because many cross-library failures occur when two individually reasonable local policies violate each other's assumptions.

\para{Controls} Controls are knobs, APIs, environment variables, and command-line flags that change a component attribute or a resource budget.

\para{Signals} Signals describe what to measure and what the measurement indicates. The \texttt{probe} field is the string name of a virtual probe that will be resolved at runtime to a profiling script from \system's profiling toolbox, which is discussed in \Cref{sec:implementation}.
Examples include active thread count, context-switch rate, CPU utilization, GPU memory use, and cgroup memory events.
Signals allow static contract composition to be checked against runtime evidence. 
A hypothesis is not considered active merely because two contracts mention the same resource; \system collects the relevant signals before spending trials on candidate actions. 

\subsection{\lpcl}
\label{sec:lpcl}

\begin{figure*}[t]
\centering
\small
\resizebox{0.97\textwidth}{!}{%
\begin{tabular}{rcl}
\(K\) & ::= &
\texttt{contract} \(Name\) \(\{\) \(D^*\) \(\}\) \\[1mm]

\(D\) & ::= &
\(Component \mid Use \mid Assume \mid Control \mid Signal\) \\[1mm]

\(Component\) & ::= &
\texttt{component} \(Name\) \texttt{:} \(Kind\) \(\{\)
\texttt{evidence=} \(Evidence\)\texttt{;}
\(\}\) \\[1mm]

\(Use\) & ::= &
\texttt{use} \(Name\) \(\{\)
\texttt{by=} \(Component\)\texttt{;}
\texttt{resource=} \(Resource\)\texttt{;}
\(Demand\)\texttt{;}
\texttt{evidence=} \(Evidence\)\texttt{;}
\(\}\) \\[1mm]

\(Assume\) & ::= &
\texttt{assume} \(Name\) \(\{\)
\texttt{resource=} \(Resource\)\texttt{;}
\(Predicate\)\texttt{;}
\texttt{evidence=} \(Evidence\)\texttt{;}
\(\}\) \\[1mm]

\(Control\) & ::= &
\texttt{control} \(Name\) \(\{\)
\texttt{target=} \(Target\)\texttt{;}
\texttt{effect=} \(Effect\)\texttt{;}
\texttt{evidence=} \(Evidence\)\texttt{;}
\(\}\) \\[1mm]

\(Signal\) & ::= &
\texttt{signal} \(Name\) \(\{\)
\texttt{target=} \(Target\)\texttt{;}
\texttt{probe=} \(ProbeName\)\texttt{;}
\texttt{evidence=} \(Evidence\)\texttt{;}
\(\}\) \\[2mm]

\(Resource\) & ::= &
\texttt{cpu.compute} \(\mid\) \texttt{cpu.placement} \(\mid\)
\texttt{gpu.compute} \(\mid\) \texttt{gpu.memory} \(\mid\)
\texttt{gpu.bandwidth} \(\mid\) \\
&& \texttt{mem.capacity} \(\mid\) \texttt{mem.bandwidth} \(\mid\)
\texttt{io.bandwidth} \(\mid\) \texttt{io.queue} \\[1mm]

\(Kind\) & ::= &
\texttt{executor} \(\mid\) \texttt{queue} \(\mid\) \texttt{allocator} \(\mid\)
\texttt{backend} \\[1mm]

\(Demand\) & ::= &
\(\mathit{Demand}(\texttt{cpu.compute})\) \(\mid\)
\(\mathit{Demand}(\texttt{cpu.placement})\) \(\mid\)  \\
&& \(\mathit{Demand}(\texttt{gpu.compute})\) \(\mid\)
\(\mathit{Demand}(\texttt{gpu.memory})\) \(\mid\)
\(\mathit{Demand}(\texttt{gpu.bandwidth})\) \(\mid\) \\
&& \(\mathit{Demand}(\texttt{mem.capacity})\) \(\mid\)
\(\mathit{Demand}(\texttt{mem.bandwidth})\) \(\mid\)
\(\mathit{Demand}(\texttt{io.bandwidth})\) \(\mid\)
\(\mathit{Demand}(\texttt{io.queue})\) \\[1mm]

\(\mathit{Demand}(\texttt{cpu.compute})\) & ::= &
\texttt{workers in} \([Expr, Expr]\) \(\mid\)
\texttt{workers=} \(Expr\) \\

\(\mathit{Demand}(\texttt{cpu.placement})\) & ::= &
\texttt{cores subset\_of} \(CpuSet\) \(\mid\)
\texttt{unconstrained} \\

\(\mathit{Demand}(\texttt{gpu.compute})\) & ::= &
\texttt{devices in} \([Expr, Expr]\) \\

\(\mathit{Demand}(\texttt{gpu.memory})\) & ::= &
\texttt{bytes in} \([Expr, Expr]\) \\

\(\mathit{Demand}(\texttt{gpu.bandwidth})\) & ::= &
\texttt{bandwidth=} \(Expr\) \\

\(\mathit{Demand}(\texttt{mem.capacity})\) & ::= &
\texttt{bytes in} \([Expr, Expr]\) \\

\(\mathit{Demand}(\texttt{mem.bandwidth})\) & ::= &
\texttt{bandwidth=} \(Expr\) \\

\(\mathit{Demand}(\texttt{io.bandwidth})\) & ::= &
\texttt{bandwidth=} \(Expr\) \\

\(\mathit{Demand}(\texttt{io.queue})\) & ::= &
\texttt{depth in} \([Expr, Expr]\) \\[1mm]

\(Predicate\) & ::= &
\texttt{not\_nested} \(\mid\)
\texttt{exclusive} \\[1mm]

\(Target\) & ::= &
\(Component\) \(\mid\)
\(Use\) \(\mid\)
\(Resource\) \\[1mm]

\(Effect\) & ::= &
\texttt{set} \(Target\) \texttt{to} \(Expr\) \(\mid\)
\texttt{decrease} \(Target\) \(\mid\)
\texttt{partition} \(Resource\) \(\mid\)
\texttt{disable} \(Target\)
\end{tabular}%
}
\caption{
LPCL context-free grammar.
The grammar contains the five declaration types used in this paper:
\texttt{component}, \texttt{use}, \texttt{assume}, \texttt{control}, and \texttt{signal}.
For \texttt{signal}, the \texttt{probe} field names a virtual probe in \system's profiling toolbox.
The resource-specific demand forms define how each resource type expresses demand.
The \texttt{evidence} field records raw evidence related to this declaration for user auditing purposes and is not used by the runtime.
}
\label{fig:lpcl}
\end{figure*}

\Cref{fig:lpcl} shows the context-free grammar of \lpcl. 
A contract begins with a named \texttt{contract} block and contains a sequence of declarations. 
The declaration types correspond to the contract concepts above: \texttt{component}, \texttt{use}, \texttt{assume}, \texttt{control}, \texttt{signal}. 
This structure is intentionally small. 
The goal is not to encode every detail of a library implementation, but to capture enough typed information for \system to generate and validate hypotheses about cross-library resource contention.

\lpcl has a fixed resource vocabulary. 
A \texttt{use} declaration must refer to one of the shared contention resources in \Cref{tab:resources}. 
\lpcl deliberately does not allow arbitrary resource names. 
For example, \texttt{cpu.thread}, \texttt{worker.pool}, and \texttt{gpu.stream} are rejected by the \lpcl checker because they are not part of the resource vocabulary.
This shared resource vocabulary allows contracts from different libraries to compose: two libraries may use different internal mechanisms, but they can still be compared through the same resource vocabulary.

The \texttt{Demand} fields are resource-specific: CPU compute uses worker counts, CPU placement uses core sets, GPU memory uses bytes, and I/O queue resources use queue depth.

A control has a target and an effect, such as setting a worker count, decreasing a queue depth, partitioning a resource, or disabling a cache. 
A signal has a target and a probe.
For example, a thread count signal may indicate \texttt{cpu.compute} pressure.

Finally, \lpcl includes a metadata field \texttt{evidence}. 
The \texttt{evidence} field records where the declaration came from, such as a documentation page, API reference, or source-code fragment. It has no effect on composition or autotuning. 
This field is especially important for LLM-assisted contract generation because it makes generated contracts checkable and auditable rather than free-form advice.

\begin{table}[t]
\centering
\small
\begin{tabular}{p{0.31\linewidth}p{0.58\linewidth}}
\toprule
\textbf{Resource} & \textbf{Meaning} \\
\midrule
\texttt{cpu.compute} & CPU execution capacity consumed by runnable workers, thread pools, and task executors. \\
\texttt{cpu.placement} & Affinity and core-placement constraints over CPU sets and NUMA domains. \\
\texttt{gpu.compute} & GPU compute units consumed by kernels. \\
\texttt{gpu.memory} & GPU memory capacity consumed by buffers. \\
\texttt{gpu.bandwidth} & GPU memory traffic throughput from reads and writes. \\
\texttt{mem.capacity} & Host memory capacity consumed by buffers. \\
\texttt{mem.bandwidth} & Host memory traffic throughput from reads and writes. \\
\texttt{io.bandwidth} & Storage or network throughput. \\
\texttt{io.queue} & Outstanding I/O concurrency or request queue depth. \\
\bottomrule
\end{tabular}
\caption{The \lpcl resource vocabulary. The language can be extended with new resource vocabulary, but the initial set shown here focuses on common shared resources that can lead to cross-library contention.}
\label{tab:resources}
\end{table}

\subsection{Contract Composition and Hypothesis Generation}

Given contract instances \(C_1,\ldots,C_n\), \system builds a resource ledger. Each ledger entry records a component, resource, demand interval, assumptions, controls, and signals. The composition algorithm applies a small set of conservative rules. Each rule emits a conflict hypothesis.

\para{Budget violation} The sum of upper-bound demands can exceed a known budget. For CPU workers, this rule detects the common case where two libraries can each create up to \texttt{host.cpu\_slots} workers, which may lead to oversubscription.

\para{Assumption violation} One contract states \texttt{not\_nested}, \texttt{exclusive}, or \texttt{competitors at\_most 0} for a resource and another contract uses that resource in an overlapping phase.

\para{Placement conflict} Two entries constrain \texttt{cpu.placement} to overlapping CPU sets. Placement conflicts can also occur when one library pins workers and another assumes unconstrained scheduling.

For example, consider a workload that uses both OpenMP and OpenBLAS. These two libraries expose different controls and have different internal executor components, but both components consume \texttt{cpu.compute}. 
By mapping both uses into the same resource ledger, \system can derive a reusable oversubscription hypothesis. The generated hypothesis will specify that \system should collect context-switch rate and CPU utilization. It may list related controls such as changing an environment variable that controls the BLAS thread count.

\subsection{Hypothesis-guided Autotuning}

\system's runtime loop has four phases.

\para{Baseline profiling} The runtime launches the target command, profiles the process tree, collects signals from all startup hypothesis profiles, and records the objective metric. The default objective is wall time, but the user may specify a workload metric such as throughput.

\para{Hypothesis validation} For each hypothesis, \system collects observed signals. For example, a CPU oversubscription hypothesis may require a high runnable-thread count and a high context-switch rate as signals. If the signals are present, the hypothesis is validated.

\para{Candidate generation} Only validated hypotheses produce candidate configurations. \lpcl \texttt{control} declarations define the legal values for each control. A candidate may coordinate multiple controls associated with one validated conflict. For example, an oversubscription candidate may jointly limit the inner OpenMP and OpenBLAS thread pools while adjusting outer concurrency.

\para{Execution and selection} Candidate configurations run through the same profiled executor as the baseline. \system accepts a candidate only if the configured objective improves. When multiple hypotheses are validated, \system tries candidates from each hypothesis in a round-robin manner and stops when the given tuning budget is exhausted, the objective converges, or no more candidates are available.

\section{Implementation}
\label{sec:implementation}

We implemented a prototype of \system as a generic end-to-end autotuning framework. The implementation has three layers: \lpcl generation, contract composition, and hypothesis-guided autotuning.

\subsection{Automatic \lpcl Generation}

\lpcl can be written by experts, but the prototype also includes an LLM-assisted generation pipeline to reduce human effort. This also makes \system an end-to-end autotuning framework that does not require users to be systems or library experts. Generation is bounded by schemas and has four passes.

\para{Ingestion and chunking} The generator searches for relevant information online and ingests local documentation, README files, source-code snippets, and web pages. It chunks text into small segments and indexes them by source IDs.

\para{Evidence extraction} An extraction prompt asks the LLM to emit evidence records for components, resource uses, controls, signals, and assumptions. Each evidence item must cite source chunks and use only the fixed resource vocabulary. Unsupported statements are omitted rather than guessed.

\para{Synthesis} A second prompt asks the LLM to synthesize an \lpcl contract from evidence records. The prompt includes the \lpcl schema and extracted evidence. The LLM must choose which evidence to include, how to map it to the schema, and how to resolve conflicts. The output is a YAML intermediate representation that can be checked and rendered to \lpcl via a deterministic compiler.

\para{Repair and compilation} The LLM-generated \lpcl will be checked against the language syntax and grammar. If validation fails, a bounded repair loop rewrites only invalid sections. Users can also manually check evidence blocks to validate that the \lpcl generated by the LLM matches the facts. In our experience, the LLM is able to produce a valid \lpcl in one run without human correction using recent LLM models such as GPT-5.4.

This pipeline uses LLMs where natural-language understanding is useful, but all runtime artifacts are structured and checked. At tuning time, \system operates on validated contracts and hypotheses rather than raw LLM outputs.

\subsection{Virtual Probes and Profiled Execution}

Profiling is the step that turns a statically generated conflict hypothesis into workload-specific evidence. However, \lpcl does not bind a signal directly to a profiling script. A direct binding would make contracts platform-specific: the script used to measure a context-switch rate on Linux/x86 may differ from the script used on another operating system, processor, accelerator stack, or cluster environment. It would also duplicate information across contracts, since different libraries may rely on the same runtime signal. For example, an \openblas contract and a \libtorch contract may both need to observe thread pressure or context-switch behavior when validating a \texttt{cpu.compute} contention hypothesis.

To keep contracts independent from profiling implementations, \system uses \emph{virtual probes}. A virtual probe is a symbolic probe name. For example, a signal may request the virtual probes \texttt{context\_switch\_rate} and \texttt{threads\_count}. During hypothesis generation, \system resolves each virtual probe name to a concrete collector in the \system profiling toolbox. The toolbox contains common platform-specific collectors shipped with \system, as well as user-provided or LLM-generated collectors. For example, on a Linux/x86 platform, the virtual probe \texttt{context\_switch\_rate} may be resolved to \texttt{context\_switch\_rate\_linux\_x86.sh}. If multiple contracts request the same virtual probe, they are mapped to the same collector and collected once for the run.

This indirection separates three concerns. The contract specifies what signal is useful for validating a library-level hypothesis. The profiling toolbox specifies how that signal is collected on a particular platform. The runtime decides whether to invoke each collector based on the active hypotheses for the current workload. As a result, the same \lpcl contract can be reused across platforms, and the same profiling script can serve many libraries and hypotheses.

The profiled executor launches the workload command and starts the signal collectors required by the active hypothesis profiles. Probes can be one-shot or session-based. A one-shot probe records a snapshot, such as an affinity mask or a device-memory summary. A session-based probe runs alongside the workload and samples periodically, such as collecting thread counts, context-switch rates, GPU utilization, or I/O throughput over time. The executor records all measurements in a single run artifact so that baseline and candidate executions are comparable.

After the baseline run, \system uses the collected probe outputs to validate hypotheses. Only hypotheses supported by the requested probe evidence generate candidate configurations. The same profiled executor is then used for candidate runs, allowing \system to compare both the objective metric and the validation signals before accepting a configuration.

\section{Evaluation}
\label{sec:evaluation}

All experiments are performed on a machine with two 12-core Intel E5-2650 v4 sockets, 128 GB of RAM, and 4 NVIDIA Tesla P100-SXM2 GPUs,
running Ubuntu 22.04 with kernel 5.15.0. The LLM model used in this evaluation is GPT-5.4 with medium thinking effort.

\subsection{LibContentionBench}
\label{sec:eval-LibContentionBench}

Evaluating cross-library contention requires a benchmark whose unit of evaluation is a \emph{library interaction}, not a single kernel, algorithm, DBMS, or monolithic application. Existing benchmark suites cover important adjacent goals, but they are not organized around this problem. Architecture and heterogeneous-computing suites such as PARSEC~\cite{parsec_benchmark} and Rodinia~\cite{rodinia_benchmark} study applications and kernels with diverse hardware behavior, but they do not expose independently managed libraries and their resource controls as first-class tuning targets. End-to-end ML, database, cloud, and latency-critical benchmarks such as MLPerf~\cite{mlperf_training}, TPC-H~\cite{tpch}, CloudSuite~\cite{cloudsuite}, and TailBench~\cite{tailbench} provide realistic workload drivers and standardized metrics. These suites are valuable for comparing complete systems, but they do not make independently managed library runtimes, their resource assumptions, and their cross-library conflicts the object of evaluation. Autotuning and algorithm-configuration suites such as AClib~\cite{aclib} and HPOBench~\cite{hpobench} provide standardized parameterized objectives for comparing optimizers. They are closer to the evaluation methodology we need, but their abstraction is a target algorithm or HPO objective function, not a workflow composed of library runtimes with independent resource assumptions.

\begin{figure*}[t]
  \centering
  \includegraphics[width=0.93\textwidth]{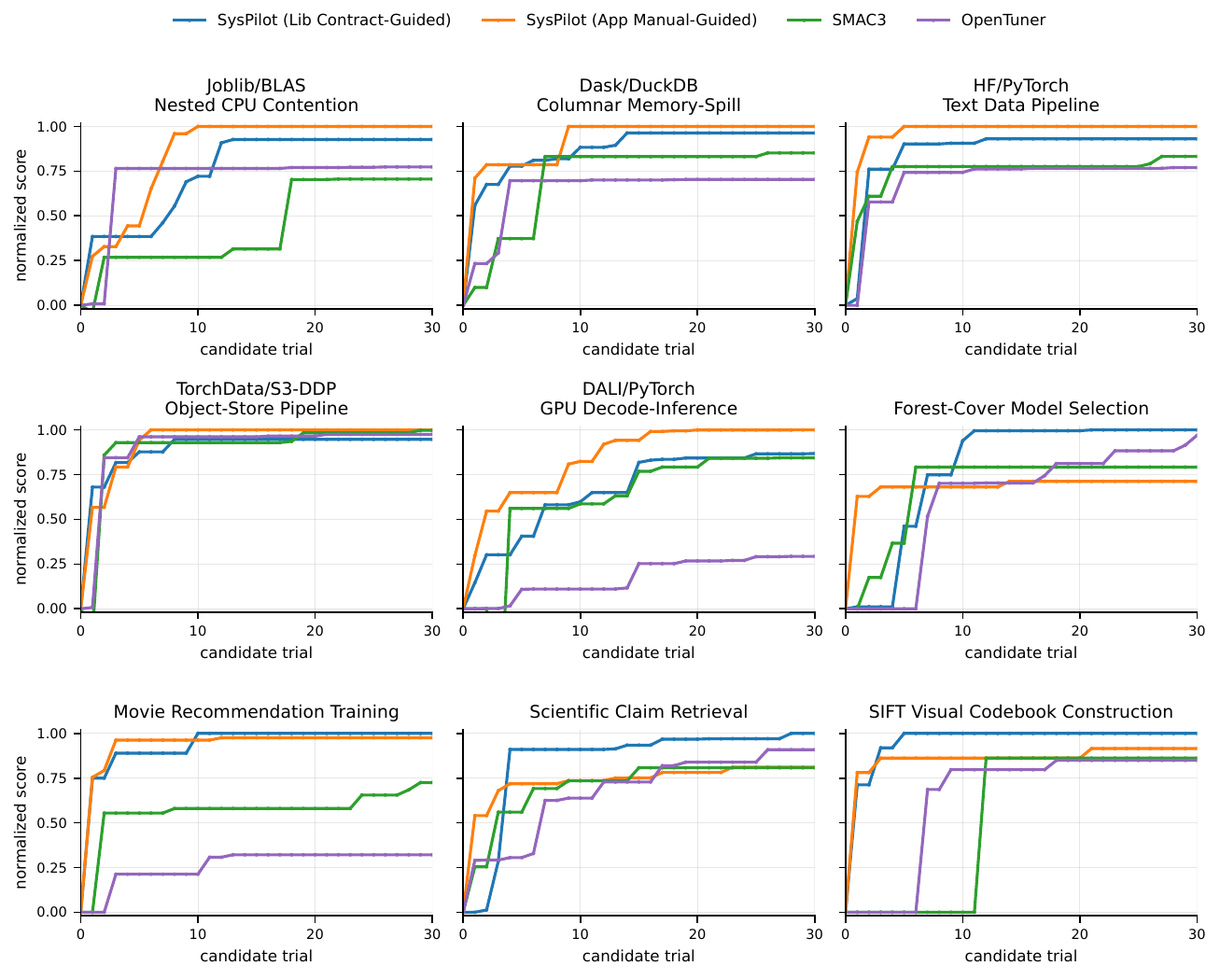}
  \caption{Best-so-far normalized score over 30 candidate trials for the nine benchmarks in LibContentionBench. The score is normalized with respect to the baseline and the best observed reference across the four methods. \system with library contracts converges faster than generic autotuners and reaches near-oracle quality within a small budget. (Upper left is better)}
  \label{fig:LibContentionBench}
\end{figure*}

We built LibContentionBench to fill this gap. Its nine benchmarks compose widely used ML, data-processing, and HPC libraries and expose their library and runtime configuration knobs. It includes five synthetic-data applications, including Joblib/BLAS Nested CPU Contention, Dask/DuckDB Columnar Memory-Spill, Hugging Face / PyTorch Text Data Pipeline, TorchData/S3-DDP Object-Store Pipeline, and DALI / PyTorch GPU Decode-Inference, whose input data is generated and whose workload sizes are bounded so that repeated tuning runs are reproducible and affordable. It also includes four real-data applications: Forest-Cover Model Selection, Movie Recommendation Training, Scientific Claim Retrieval, and SIFT Visual Codebook Construction, to demonstrate the performance of \system on real-world workloads.

At the same time, each workflow preserves a realistic contention pattern in which independently managed library runtimes compete for CPU cores, memory capacity, memory bandwidth, I/O bandwidth, object-store access, or GPU resources. Input data, workload sizes, and algorithmic settings remain fixed during each tuning session; only the exposed resource controls may change.

LibContentionBench is not intended to be exhaustive. Instead, it is designed to cover the main resource and control dimensions of cross-library contention. The nine workflows span CPU oversubscription, NUMA-sensitive CPU partitioning, host-memory and spill-path contention, data-pipeline worker and queue contention, object-store input contention, distributed-training input coordination, and GPU decode/inference co-scheduling. This coverage lets us evaluate whether a tuning method can reason across CPU, memory, I/O, and GPU resources, rather than focusing on only one resource type or one familiar library. Appendix~A provides a detailed description of each application.

\subsection{Application-Level Manual Baseline}
\label{sec:eval-appmanual}

All tuners receive the same target command, legal tuning surface, primary objective, timeout policy, and correctness constraints. The known optimal configuration (oracle) is hidden from tuners and used only to normalize results. The primary budget is 30 candidate configuration trials; we report best-so-far quality after each candidate execution.

Because workflows have different baseline-to-oracle headroom, we normalize performance by the oracle gap. Let \(S^{base}_w\) be the baseline score for workflow \(w\), \(S^{oracle}_w\) be the oracle score, and \(S^{best}_{m,w}(b)\) be method \(m\)'s best score after \(b\) candidate executions. We report
\begin{equation}
  Q_{m,w}(b) =
  \frac{S^{best}_{m,w}(b)-S^{base}_w}
       {S^{oracle}_w-S^{base}_w}.
\label{eq:normalized-gap}
\end{equation}
For throughput and fair-share metrics, \(S\) is the measured score. For lower-is-better runtime metrics, we first convert runtime to a speedup-style score relative to the baseline. Thus \(Q=0\) is baseline quality and \(Q=1\) is oracle quality.

We compare four methods. \system (Lib Contract-Guided) uses reusable library contracts and the runtime loop described in \Cref{sec:design}. \system (App Manual-Guided) uses the same executor and candidate-validation machinery, but replaces reusable contracts with a per-workflow application manual that describes the application-specific information. This represents a strong application-aware baseline that adopts the idea of the application-specific LLM-assisted autotuners in database and storage systems \cite{dbbert,gptuner,lambda_tune} while ensuring fairness by using the same \system tuning pipeline and the same LLM model.
SMAC3 and OpenTuner receive the same legal search space and objective but no library contracts~\cite{smac,opentuner}.

\subsection{End-to-end Tuning Results}
\label{sec:eval-main-results}

\begin{table}[t]
  \centering
  \small
  \setlength{\tabcolsep}{4pt}
  \begin{tabular}{lcccc}
    \toprule
    \multirow{2}{*}{Tuner}
      & \multicolumn{2}{c}{Trials to target $\downarrow$}
      & \multicolumn{2}{c}{Avg. score $\uparrow$} \\
    \cmidrule(lr){2-3}
    \cmidrule(lr){4-5}
      & $@0.8$ & $@0.9$ & $@10$ runs & $@30$ runs \\
    \midrule
    \makecell[l]{\system\\(App Manual-Guided)}
      & 6 & 12 & 0.896 & 0.935 \\
    \makecell[l]{\system\\(Lib Contract-Guided)}
      & 8 & 12 & 0.879 & 0.960 \\
    SMAC3
      & 24 & $>30$ & 0.611 & 0.824 \\
    OpenTuner
      & $>30$ & $>30$ & 0.625 & 0.729 \\
    \bottomrule
  \end{tabular}
  \caption{Aggregate convergence summary across the nine LibContentionBench benchmarks. Trials to target columns (lower is better) report the first candidate execution at which the average normalized score reaches the target; $>30$ means the target is not reached within the 30-trial budget. Average score columns (higher is better) report the average normalized best-so-far score at fixed budgets.}
  \label{tab:LibContentionBench-summary}
\end{table}

\Cref{fig:LibContentionBench} shows the best-so-far normalized score over the tuning budget. \system with library contracts reaches the 0.8 average target in 8 trials and the 0.9 target in 12 trials. The application-manual variant reaches these targets in 6 and 12 trials, respectively. Among the generic autotuners, SMAC3 reaches 0.8 only after 24 trials, while OpenTuner does not reach it within the budget; neither generic method reaches 0.9.

At 30 trials, contract-guided tuning achieves the highest average normalized score, 0.960, compared with 0.935 for the application-manual baseline, 0.824 for SMAC3, and 0.729 for OpenTuner (\Cref{tab:LibContentionBench-summary}). Application manuals provide strong target-specific guidance, particularly for the synthetic-data workflows. These workflows have intentionally bounded workload structures and relatively simple, well-defined library interactions, making it feasible to write a manual that accurately captures the important resource usage patterns and tuning opportunities.

This advantage becomes less consistent on the real-data applications, whose execution involves more complex interactions among libraries and runtime components. In such applications, it is difficult for a manually written description to capture every performance-relevant detail in advance, and omitted details can make the guidance less effective. Library contracts instead describe reusable performance behavior at the library level and compose this information for the application being tuned. As a result, \system retains the benefits of structured performance knowledge while avoiding the need to construct a complete application-specific description for every workload. This allows contract-guided tuning to achieve comparable or better aggregate performance than application-manual guidance while substantially outperforming generic autotuners. Both of them outperform generic autotuners in terms of convergence speed and final quality.

\subsection{Case Study: How \system Narrows the Search Space}
\label{sec:eval-nested-cpu-blas}

This section uses the Joblib/BLAS nested CPU contention benchmark as a
case study to explain why \system outperforms generic autotuners under the same 30-trial budget. The workload contains nested parallelism: Joblib launches multiple outer tasks, and each Joblib task performs NumPy matrix multiplication. NumPy dispatches this computationally intensive operation to the numerical runtime linked into its build, such as OpenBLAS, MKL, or BLIS. Depending on the
build, additional thread-pool controls such as OpenMP may also apply. A configuration $(j,o,b,m,l,k)$ specifies Joblib workers, OpenMP, OpenBLAS, MKL, and BLIS thread limits, and the Joblib backend. Numeric controls range over $[1,48]$ and the backend has two choices, giving $2\times48^5=509{,}607{,}936$ requested tuples.

The executed NumPy build uses OpenBLAS. Runtime evidence identifies $o$, $m$, and $l$ as inactive, so \system fixes them to one and searches $(j,1,b,1,1,k)$: $48\times48\times2=4{,}608$ tuples. Contracts then relate outer Joblib workers and inner BLAS capacity to the shared CPU resource. Generic tuners lack this applicability information and must use their observations both to identify influential controls and to optimize their values. Moreover, many configurations within this reduced space can be eliminated because their combinations of outer workers and inner BLAS threads oversubscribe the 48 physical cores. The reduction therefore removes two major sources of wasted trials.

Within 30 trials, contracts reach a normalized score of 0.927, compared with 0.706 for SMAC3 and 0.774 for OpenTuner; the application manual reaches 1.0. The distinction is how the budget is spent: \system focuses on Joblib concurrency, the active OpenBLAS pool, and the backend, rather than repeatedly varying controls that do not affect this execution.

\subsection{Cross-application Manual Transfer}
\label{sec:eval-manual-transfer}

The previous experiment compares \system with application-level manuals that are written specifically for each workflow. Such manuals are useful as target-specific priors, but they do not by themselves demonstrate reusable performance knowledge. To evaluate whether an application manual can be reused as a tuning artifact, we run a cross-application manual-transfer experiment on three PyTorch-related workflows: the Hugging Face / PyTorch Text Data Pipeline, the TorchData/S3-DDP Object-Store Pipeline, and the DALI/PyTorch GPU Decode-Inference workflow.

All three workflows contain PyTorch, but they compose PyTorch with different surrounding libraries and stress different contention resources. The text pipeline combines contention among dataset workers, PyTorch CPU threads, and DataLoader workers. The object-store pipeline involves contention over object-store I/O and PyTorch CPU threads. The GPU pipeline involves contention over shared GPU compute units and GPU memory. This makes the set a useful test of whether an application-level manual transfers across applications that share a major library family but differ in cross-library contention structure.

For each target workflow, every method receives the target command, objective, legal knob schema, and correctness constraints. The only transferred artifact is the natural-language application manual. A diagonal entry uses the target's own manual and represents the fresh-manual baseline. An off-diagonal entry uses a manual written for a different workflow without editing it for the target. We measure the number of candidate executions required to reach at least 98\% of the oracle score. Lower values are better.

\begin{figure}[t]
  \centering
  \includegraphics[width=\linewidth]{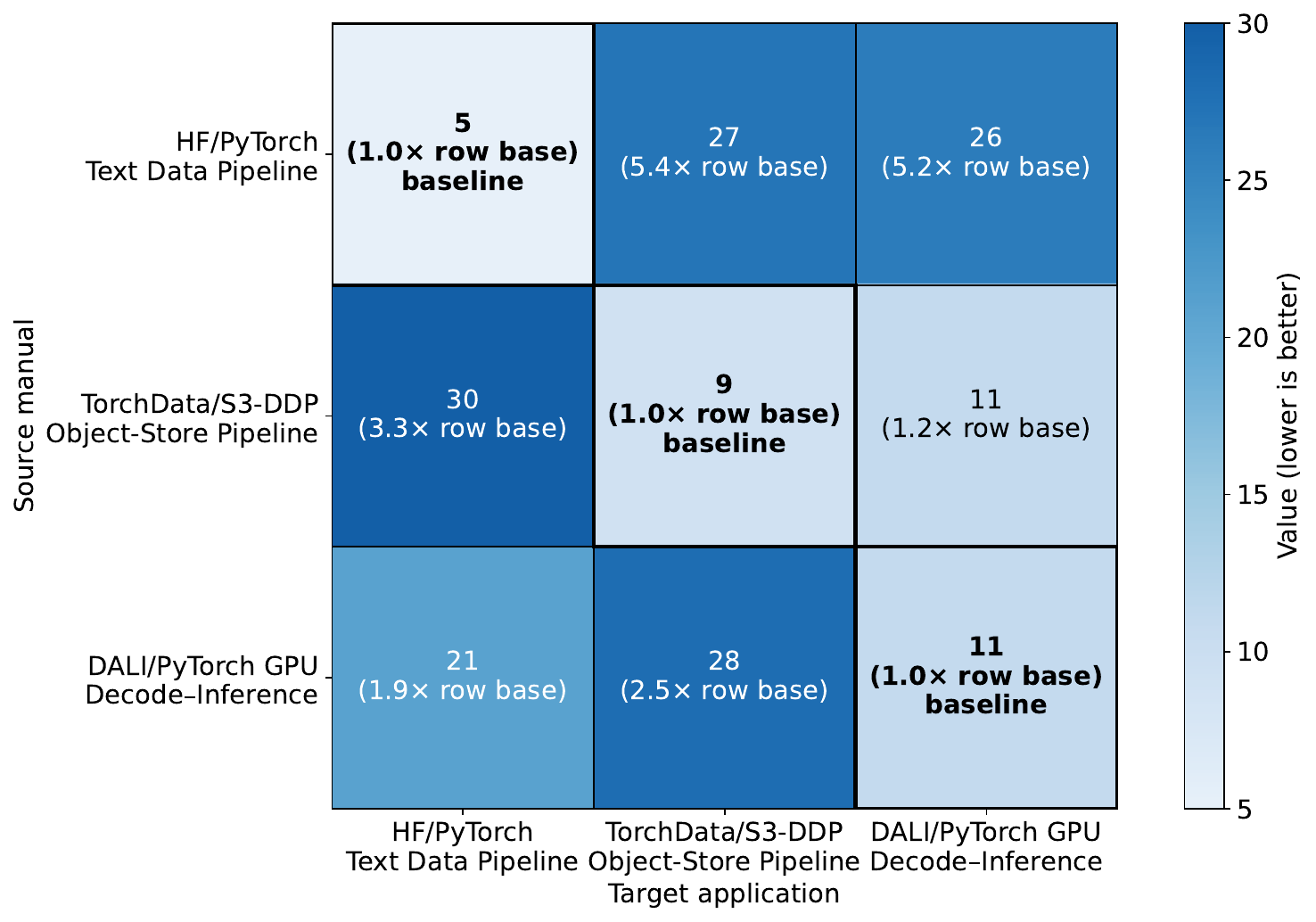}
  \caption{Cross-application manual transfer among three PyTorch-related workflows. Rows are source manuals and columns are target applications. Each cell reports the number of trials required to reach 98\% of the oracle score. Diagonal cells use the target's own manual and serve as the row baseline; off-diagonal cells reuse a different application's manual without editing it. Values in parentheses show slowdown relative to the diagonal baseline for the same source manual. (Lighter color is better)}
  \label{fig:manual-transfer}
\end{figure}

\Cref{fig:manual-transfer} shows that application manuals are effective when written for the target but brittle when reused. With the matching manual, the HF/PyTorch text pipeline reaches the 98\% threshold in 5 trials, the TorchData/S3-DDP pipeline in 9 trials, and the DALI/PyTorch GPU pipeline in 11 trials. Off-diagonal transfer is much less reliable: the HF/PyTorch manual requires 27 trials on the TorchData/S3-DDP target and 26 trials on the DALI/PyTorch target, corresponding to 5.4$\times$ and 5.2$\times$ slowdowns relative to its own fresh-manual baseline. The TorchData/S3-DDP manual requires 30 trials on the HF/PyTorch target, a 3.3$\times$ slowdown. The DALI/PyTorch manual requires 21 and 28 trials on the other two targets, corresponding to 1.9$\times$ and 2.5$\times$ slowdowns.

One transfer case remains close to its source baseline: using the TorchData/S3-DDP manual for the DALI/PyTorch target reaches the threshold in 11 trials, only 1.2$\times$ the TorchData manual's own baseline. This exception is important because it shows that transfer is not forced to fail. Rather, it is unstable. A whole-application manual entangles stable library behavior, application-specific structure, and deployment-specific assumptions. Reusing it transfers all three together, including assumptions that may not hold for the new target. \system instead factors the stable part into library contracts and composes only the contracts needed for the target workflow.

These results support the design choice in \system. Application manuals entangle three kinds of knowledge: stable library behavior, application-specific call structure, and deployment-specific contention assumptions. Reusing the entire manual transfers all three together, including assumptions that may not hold for the new target. \system instead factors the stable part into library-level contracts. Contracts for PyTorch, DALI, fsspec/S3, DataLoader, or DDP can be selected and composed for the target workflow, while profiling validates which resource-conflict hypotheses are active in the current run. Thus, \system does not require a fresh application manual for each new workflow, and it avoids treating a previous application's manual as a monolithic tuning prior.

Overall, the manual-transfer experiment shows that fresh application manuals are strong target-specific baselines, but they are poor reusable artifacts. Even within a small set of PyTorch-related applications, transferring a manual from one workflow to another can increase the trials needed to reach near-oracle performance by up to 5.4$\times$. This motivates using composable library-level contracts as the reusable unit of tuning knowledge instead of whole-application manuals.

\subsection{Ablation Study} 
\label{sec:ablation-study}

We next investigate how different \system features contribute to search-efficiency improvement.
This ablation uses the same transformer hyperparameter search workload from the motivating example in \Cref{sec:motivation}: two concurrent C++ \libtorch transformer tasks run on a 24-core machine with two different hyperparameter settings, and the tuner must find the core allocation that maximizes performance.

We therefore compare four variants that differ in how much structured information is available to the tuner. 
The primary metric is the number of candidate trials required to find the oracle configuration; lower is better. 
For the knob-only baseline, the tuner is capped at 30 trials, so we report 30+ when it fails to find the oracle within the budget.

\begin{table}[t]
  \centering
  \small
  \caption{
    Ablation study on two concurrent C++ \libtorch transformer tasks. (Lower is better).
  }
  \label{tab:ablation-transformer}
  \begin{tabular}{lc}
    \toprule
    Variant & Trials to oracle \\
    \midrule
    Full \system & 7 \\
    Without probes/profiling & 10 \\
    Raw documentation instead of \lpcl & 28 \\
    Knob-only autotuner & 30+ \\
    \bottomrule
  \end{tabular}
\end{table}

Table~\ref{tab:ablation-transformer} shows that each source of structure contributes to search efficiency. 

\para{The full system} The full system uses the complete \system pipeline with all features described in the paper enabled. It finds the oracle in 7 trials because \lpcl identifies the relevant shared resource and candidate interventions, while profiling validates which contract-derived hypotheses are active in the current run.

\para{No Profiling} This variant keeps the \lpcl contracts and the contract-composition logic, but removes virtual probes and runtime profiling. As a result, \system can still generate hypotheses from the contracts, but it cannot check which hypotheses are actually active in the current workload. The tuner therefore treats all contract-derived hypotheses as plausible and generates candidates for a larger set of possible conflicts. This increases the number of trials to 10. The result shows that \lpcl alone provides useful structure, but profiling further reduces the search space by filtering out hypotheses that are not supported by runtime evidence.

\para{Raw Documentation without \lpcl} This variant replaces \lpcl contracts with raw natural-language documentation about \libtorch, while keeping the same tuning harness and knob schema. The tuner can read relevant text, but the information is not represented as typed resources, controls, signals, and assumptions. Therefore, the system cannot deterministically compose contracts, build a resource ledger, or translate library knowledge into a small set of resource-conflict hypotheses. Candidate generation relies on free-form interpretation of the documentation, which is much less precise. This variant requires 28 trials to find the oracle. The result shows that the benefit is not simply from giving the tuner more text; the information must be structured so that it can guide hypothesis generation and narrow the root cause of contention.

\para{Knob-only autotuner baseline} This baseline receives only the legal tuning-knob schema, the objective, and the same execution harness. It receives no \lpcl contracts, no raw documentation, and no library-level explanation of what the knobs control. Consequently, the tuner has no information about which libraries are competing, which resource is shared, or which actions are likely to relieve contention. The search degenerates into black-box exploration over a large combinatorial configuration space. This baseline does not find the oracle within the 30-trial budget. The result shows that the improvement does not come from the autotuning framework alone; it comes from the library-level information supplied through \lpcl and refined through profiling.

Overall, the ablation study shows that the improvement does not come from the autotuning framework alone. The knob-only baseline uses the same tuning interface but receives no library-level information, and it fails to find the oracle within the trial budget. Adding \lpcl changes the problem from black-box knob exploration into hypothesis-guided tuning: the contract identifies the relevant resources, controls, and contention mechanisms, allowing \system to generate a small set of candidate actions targeted at the likely root cause. The raw-documentation variant shows that simply giving the tuner more text is not enough. \lpcl improves over raw \libtorch documentation because it represents library knowledge in a structured, composable form that can be checked, combined, and translated into resource-conflict hypotheses. Profiling then provides a second reduction in search space by validating which contract-derived hypotheses are active in the current workload, preventing the tuner from spending trials on plausible but irrelevant interventions.

\section{Limitations}
\label{sec:limitations}

\para{Contract quality} \system is only as good as its contracts. The quality of LLM-generated contracts depends largely on the available documentation. Documentation may omit performance behavior, source code may be hard to interpret, and an LLM extractor can miss or misclassify evidence. 

\para{Static false positives and false negatives} Contract composition is conservative. It may flag resources that are never used concurrently, or miss interference caused by an undocumented resource. Profiling is intended to filter false positives, but false negatives require better contracts or broader probes.

\para{Version and build sensitivity} Libraries change behavior across versions, build options, and backends. The current prototype records evidence but does not yet provide full version-aware retrieval, versioned compatibility checks, or automatic contract aging.

\section{Future Work}
\label{sec:future}

Several extensions follow naturally.

\para{Version-aware contracts} Future contracts should include version ranges, backend conditions, build flags, and source provenance so the store can select the right contract for a particular deployment.

\para{Distributed workflow tuning} LPCL intentionally uses a small vocabulary, and the main focus is on single-node tuning. However, the same principles apply to distributed jobs, where contracts can describe cluster resources, network links, and inter-node controls. Future work can extend the language and prototype to support distributed workflows.

\section{Conclusion}
\label{sec:conclusion}
Modern applications increasingly compose independently managed libraries, making shared-resource contention a cross-library problem rather than a single-runtime tuning problem. This paper presented \system, a cross-library autotuning framework that uses Library Performance Contracts to make library-level performance knowledge reusable. We designed \lpcl{}, a specification language that captures a library's resource demands, assumptions, controls, and runtime signals. \system synthesizes and checks these contracts, composes them into resource-conflict hypotheses, validates the hypotheses through profiling, and uses them to generate candidate configurations.

We evaluated \system on the nine benchmarks in LibContentionBench. \system with \lpcl{} finds better configurations in fewer trials than general-purpose autotuners, achieving an average normalized score of 0.960, compared with 0.824 for SMAC3 and 0.729 for OpenTuner. \system also reaches the 0.8 normalized score threshold in 3$\times$ fewer trials.

\bibliographystyle{ACM-Reference-Format}
\bibliography{references}

\clearpage
\appendix
\section{Benchmark Details}
\label{app:workloads}

This appendix describes the nine LibContentionBench workloads. Input data, workload sizes, and model and algorithm settings remain fixed during tuning; only the resource controls listed below may change. The configuration-space sizes are based on the 48 CPU slots available in our experiments and may differ on other hardware.

\begin{table*}[t]
  \centering
  \scriptsize
  \setlength{\tabcolsep}{2.5pt}
  \renewcommand{\arraystretch}{1.08}
  \begin{tabular}{@{}>{\raggedright\arraybackslash}p{0.18\textwidth}>{\raggedright\arraybackslash}p{0.22\textwidth}>{\raggedright\arraybackslash}p{0.24\textwidth}>{\raggedright\arraybackslash}p{0.20\textwidth}r@{}}
    \toprule
    Benchmark & Libraries and datasets & Shared resources & Configuration space & Best performance gain \\
    \midrule
    Joblib/BLAS Nested CPU Contention & Joblib, NumPy, OpenBLAS~\cite{numpy,openblas} & \texttt{cpu.compute} & $5.09\times10^8$ & $16.7\times$ \\
    Dask/DuckDB Columnar Memory-Spill & Dask, PyArrow, DuckDB~\cite{dask,duckdb} & \texttt{cpu.compute}, \texttt{mem.capacity}, \texttt{io.bandwidth} & $8.15\times10^9$ & $1.66\times$ \\
    HF/PyTorch Text Data Pipeline & HF Datasets, HF Tokenizers, PyTorch DataLoader~\cite{hf_datasets,pytorch_dataloader} & \texttt{cpu.compute}, \texttt{mem.capacity} & $2.17\times10^7$ & $2.35\times$ \\
    TorchData/S3-DDP Object-Store Pipeline & TorchData, fsspec/s3fs, MinIO, PyTorch DDP~\cite{pytorch} & \texttt{cpu.compute}, \texttt{mem.capacity}, \texttt{io.bandwidth}, \texttt{io.queue} & $2.57\times10^{10}$ & $1.68\times$ \\
    DALI/PyTorch GPU Decode-Inference & DALI/nvJPEG, PyTorch/cuDNN & \texttt{gpu.compute}, \texttt{gpu.memory}, \texttt{io.queue} & $6.04\times10^{12}$ & $1.84\times$ \\
    Forest-Cover Model Selection & Dask-ML, XGBoost, UCI Covertype~\cite{dask,xgboost,uci_covertype} & \texttt{cpu.compute}, \texttt{mem.capacity} & $4{,}802$ & $1.84\times$ \\
    Movie Recommendation Training & Implicit, OpenBLAS, OpenMP, MovieLens20M~\cite{openblas,openmp,movielens20m} & \texttt{cpu.compute}, \texttt{cpu.placement} & $14{,}112$ & $1.32\times$ \\
    Scientific Claim Retrieval & SentenceTransformers, PyTorch, FAISS, SciFact~\cite{sentence_transformers,pytorch,faiss,scifact} & \texttt{cpu.compute}, \texttt{mem.capacity} & $5.04\times10^9$ & $2.16\times$ \\
    SIFT Visual Codebook Construction & Spark MLlib, OpenBLAS, SIFT1M~\cite{spark,openblas,sift1m} & \texttt{cpu.compute}, \texttt{mem.capacity} & $2{,}401$ & $1.27\times$ \\
    \bottomrule
  \end{tabular}
  \caption{LibContentionBench workloads, libraries and datasets, shared resources expressed in LPCL, configuration-space sizes, and performance gains over the baseline.}
  \label{tab:workload-details}
\end{table*}

\subsection{Workload Definitions}

\para{Joblib/BLAS Nested CPU Contention}
\textbf{Input and execution.} The workload runs 800 independent tasks, each multiplying two fixed $320\times320$ matrices six times. Joblib runs the tasks in parallel, and NumPy uses the host BLAS runtime for matrix multiplication~\cite{numpy,openblas}.
\textbf{Controls.} The tunable controls are the Joblib worker count, thread limits for OpenMP, OpenBLAS, MKL, and BLIS, and the choice of a process or thread backend.
\textbf{Contention.} Joblib workers and the native numerical routines they invoke share CPU cores (\texttt{cpu.compute}). When each outer worker also uses multiple native threads, the combined demand can exceed the available CPU capacity.
\textbf{Configuration space.} The five worker and thread counts, together with two backend choices, give $2\times48^5=509{,}607{,}936$ configurations.

\para{Dask/DuckDB Columnar Memory-Spill}
\textbf{Input and execution.} Dask processes ten Parquet files, each containing 20,000 rows in row groups of 5,000. The data is read into Arrow tables and passed to DuckDB for a fixed aggregation~\cite{dask,duckdb}.
\textbf{Controls.} The tunable controls are the Dask worker count, threads per worker, worker memory limit, Arrow CPU and I/O thread counts, DuckDB memory limit, DuckDB thread count, and scratch-path layout.
\textbf{Contention.} Dask, Arrow, and DuckDB workers share CPU cores (\texttt{cpu.compute}). In-memory data and spill buffers consume host memory (\texttt{mem.capacity}), while spill operations using the same scratch path compete for I/O bandwidth (\texttt{io.bandwidth}).
\textbf{Configuration space.} The eight controls give $48^5\times4^2\times2=8.15\times10^9$ configurations.

\para{HF/PyTorch Text Data Pipeline}
\textbf{Input and execution.} Hugging Face Datasets processes 4,096 fixed text records with Tokenizers, and a PyTorch DataLoader supplies the results to a fixed training loop~\cite{hf_datasets,pytorch_dataloader}. Each record contains 32 input tokens, the maximum sequence length is 64, and the batch size is 32.
\textbf{Controls.} The tunable controls are the PyTorch thread count, dataset-mapping process count, tokenizer parallelism, DataLoader worker count, prefetch factor, and use of persistent workers.
\textbf{Contention.} Dataset mapping, tokenization, data loading, and PyTorch execution share CPU cores (\texttt{cpu.compute}). Worker processes and prefetched batches also consume host memory (\texttt{mem.capacity}).
\textbf{Configuration space.} The six controls give $48^3\times49\times2^2=2.17\times10^7$ configurations.

\para{TorchData/S3-DDP Object-Store Pipeline}
\textbf{Input and execution.} Two CPU training ranks each read 1,024 samples from four shards in a local S3-compatible object store. The workflow uses TorchData and fsspec for input requests and PyTorch DDP for training~\cite{pytorch}.
\textbf{Controls.} The tunable controls are the PyTorch thread count, S3 block size, S3 cache mode, object-listing cache policy, and DDP bucket capacity.
\textbf{Contention.} Input decoding and training share CPU cores (\texttt{cpu.compute}). Both ranks access the local object store over loopback and share its transfer bandwidth (\texttt{io.bandwidth}). Request concurrency and prefetch depth affect the number of outstanding I/O operations (\texttt{io.queue}), while caches and DDP buckets consume host memory (\texttt{mem.capacity}).
\textbf{Configuration space.} The five controls give $48\times1{,}044{,}481\times4\times2\times64=2.57\times10^{10}$ configurations.

\para{DALI/PyTorch GPU Decode-Inference}
\textbf{Input and execution.} DALI decodes a dataset of 5,000 JPEG images while PyTorch and cuDNN run convolutional inference under CUDA MPS~\cite{dali,pytorch,cudnn}. Each run has a ten-second warm-up followed by a 45-second measurement interval.
\textbf{Controls.} The tunable controls are the DALI batch size, prefetch depth, decoder load, decoder thread count, DALI MPS share, inference batch size, and PyTorch MPS share.
\textbf{Contention.} Decoding and inference kernels compete for GPU compute capacity (\texttt{gpu.compute}), and their buffers share device memory (\texttt{gpu.memory}). DALI's prefetch depth controls the number of queued batches (\texttt{io.queue}).
\textbf{Configuration space.} The six integer-valued controls give $512^2\times48^2\times100^2=6.04\times10^{12}$ discrete combinations. Decoder load is an additional continuous control.

\para{Forest-Cover Model Selection}
\textbf{Input and execution.} Dask-ML GridSearchCV evaluates six fixed model configurations using three-fold cross-validation on data from a stratified, 120,000-row UCI Covertype subset. It then refits the selected model and predicts labels for 24,000 held-out rows~\cite{dask,xgboost,uci_covertype}. The timed region includes all 18 cross-validation fits, refitting, and prediction.
\textbf{Controls.} The tunable controls are the cross-validation worker count, XGBoost native-thread count, and Dask local scheduler choice (threads or processes).
\textbf{Contention.} Concurrent cross-validation fits and their native XGBoost workers share CPU cores (\texttt{cpu.compute}). Each fit also retains training data and intermediate state, consuming host memory (\texttt{mem.capacity}).
\textbf{Configuration space.} Each integer control ranges from $0$ to $48$, and the scheduler has two choices, giving $49^2\times2=4{,}802$ configurations.

\para{Movie Recommendation Training}
\textbf{Input and execution.} Implicit trains an alternating least squares (ALS) recommendation model on all 20,000,263 MovieLens20M ratings, covering 138,493 users and 26,744 movies~\cite{movielens20m}. It uses 128 factors and 15 training iterations, then generates ten recommendations for every user.
\textbf{Controls.} The tunable controls are the ALS worker count, OpenBLAS thread count, OpenMP binding policy, and placement by CPU core or hardware thread.
\textbf{Contention.} ALS workers and BLAS kernels share CPU capacity (\texttt{cpu.compute}). Binding and placement settings affect how workers use cores and NUMA domains (\texttt{cpu.placement}).
\textbf{Configuration space.} The four controls give $49\times48\times3\times2=14{,}112$ configurations.

\para{Scientific Claim Retrieval}
\textbf{Input and execution.} SentenceTransformers encodes 5,183 SciFact abstracts and 300 queries with a fixed MiniLM model. FAISS then performs exact search and returns ten documents per query~\cite{sentence_transformers,scifact,faiss}. Timing includes worker-pool startup, corpus and query encoding, retrieval, and shutdown, but excludes loading the model in the parent process.
\textbf{Controls.} The tunable controls are the encoder process count, PyTorch thread count, tokenizer thread count, batch size, and chunk size.
\textbf{Contention.} Encoder processes and their native PyTorch and tokenizer threads share CPU cores (\texttt{cpu.compute}). Each process also holds its own model state and intermediate buffers, consuming host memory (\texttt{mem.capacity}).
\textbf{Configuration space.} The five controls give $8\times49^2\times128\times2{,}049=5.04\times10^9$ configurations.

\para{SIFT Visual Codebook Construction}
\textbf{Input and execution.} Spark MLlib builds a K-means codebook with 1,024 centers from 100,000 SIFT1M learning descriptors, each with 128 dimensions~\cite{spark,sift1m}. Each run starts a new local JVM, loads the input, trains and saves the model, assigns every descriptor to a cluster, writes the results, and shuts down.
\textbf{Controls.} The tunable controls are the CPU slots reserved per Spark task and the process-wide OpenBLAS thread limit.
\textbf{Contention.} Spark tasks and OpenBLAS kernels share CPU cores (\texttt{cpu.compute}). Concurrent Spark tasks also compete for space in the JVM heap (\texttt{mem.capacity}).
\textbf{Configuration space.} The two controls give $49^2=2{,}401$ configurations.

\end{document}